\documentclass[sigconf,nonacm]{acmart}
\AtBeginDocument{%
  }

\usepackage{booktabs}  
\usepackage{tabularx}
\usepackage{graphicx} 
\usepackage{multirow}

\begin{document}

\title{From Signal to Turn: Interactional Friction in Modular Speech-to-Speech Pipelines}

\author{Tittaya Mairittha}
\authornote{All authors contributed equally to this research}
\email{tittaya.mai@axonstech.com}
\affiliation{%
  \institution{AXONS}
  \city{Bangkok}
  \country{Thailand}
}

\author{Tanakon Sawanglok}
\email{tanakon.saw@axonstech.com}
\affiliation{%
  \institution{AXONS}
  \city{Bangkok}
  \country{Thailand}
}

\author{Panuwit Raden}
\email{panuwit.rad@axonstech.com}
\affiliation{%
  \institution{AXONS}
  \city{Bangkok}
  \country{Thailand}
}

\author{Jirapast Buntub}
\email{jirapast.bun@axonstech.com}
\affiliation{%
  \institution{AXONS}
  \city{Bangkok}
  \country{Thailand}
}

\author{Thanapat Warunee}
\email{thanapat.war@axonstech.com}
\affiliation{%
  \institution{AXONS}
  \city{Bangkok}
  \country{Thailand}
}

\author{Napat Asawachaisuvikrom}
\email{napat.asa@axonstech.com}
\affiliation{%
  \institution{AXONS}
  \city{Bangkok}
  \country{Thailand}
}

\author{Thanaphum Saiwongin}
\email{t-thanaphum.sai@axons.in.th}
\affiliation{%
  \institution{Chulalongkorn University}
  \city{Bangkok}
  \country{Thailand}
}

\begin{abstract}
While voice-based AI systems have achieved remarkable generative capabilities, their interactions often feel conversationally broken. This paper examines the interactional friction that emerges in modular Speech-to-Speech Retrieval-Augmented Generation (S2S-RAG) pipelines. By analyzing a representative production system, we move beyond simple latency metrics to identify three recurring patterns of conversational breakdown: (1) Temporal Misalignment, where system delays violate user expectations of conversational rhythm; (2) Expressive Flattening, where the loss of paralinguistic cues leads to literal, inappropriate responses; and (3) Repair Rigidity, where architectural gating prevents users from correcting errors in real-time. Through system-level analysis, we demonstrate that these friction points should not be understood as defects or failures, but as structural consequences of a modular design that prioritizes control over fluidity. We conclude that building natural spoken AI is an infrastructure design challenge, requiring a shift from optimizing isolated components to carefully choreographing the seams between them.
\end{abstract}



\renewcommand{\shortauthors}{}

\maketitle

\section{Introduction}~\label{introduction}
Spoken conversation is defined not by the mere exchange of information, but by its fluid and collaborative nature. In human dialogue, turn-taking is managed with subtle cues, interruptions are handled gracefully, and silence is a meaningful signal. While the field of Human-Computer Interaction (HCI) has long aspired to replicate this flow, current conversational AI systems often fall short. They may be linguistically intelligent—capable of generating complex, factual answers—but they remain interactionally incompetent. Their rigid turn-taking, unnatural pauses, and inability to adapt in real-time marks them immediately as machines.

The recent emergence of native, end-to-end models promises to bridge this gap, but the practical requirements of enterprise deployment—hallucination control, auditability, and access to proprietary data—ensure that modular Retrieval-Augmented Generation (RAG) architectures will remain dominant. These systems typically cascade several specialized components: Automatic Speech Recognition (ASR), retrieval mechanisms, a Large Language Model (LLM), and Text-to-Speech (TTS) synthesis.

While this modularity offers engineering control, it introduces significant interactional friction. To the user, the system appears as a single conversational partner; architecturally, however, it is a distributed network of asynchronous processes. When these components fail to coordinate perfectly, the seams show. This friction is not merely a matter of latency. A speech recognition error can cascade into a nonsensical answer; the system may become deaf to user interruptions while speaking; and the expressive nuance of the user's voice is often lost. This is particularly acute in languages lacking clear word boundaries, such as Thai, where segmentation errors propagate downstream, altering the pragmatic meaning of a user’s intent.

This paper argues that the unnaturalness of modern speech agents is not a failure of model capacity, but a structural consequence of the pipeline architecture itself. We treat the speech system not as a black box, but as an interactional infrastructure—a site where engineering trade-offs (e.g., latency vs. accuracy, control vs. fluidity) directly shape the user's experience. Through the design and analysis of a pilot S2S-RAG system, we investigate how this interplay of timing, semantics, and expressiveness defines the felt quality of the interaction. We demonstrate that achieving conversational coherence requires shifting focus from optimizing isolated modules to choreographing the complex seams between them.

\newpage
\section{Related Work}~\label{background}

Research on speech-based HCI has long established that fluent conversation relies on more than just accurate word recognition; it requires the fine-grained timing and coordination of turns \cite{schegloff2007sequence}. This fundamental norm has motivated the development of incremental dialogue architectures designed to process input and update system state turn-by-turn in real time \cite{schlangen2011incremental, skantze2021survey}. Our work builds upon this theoretical foundation, examining the practical friction that emerges when these interactional ideals are implemented within modern, high-latency neural architectures.

At the input layer, the evolution of streaming ASR has been critical for reducing the latency-accuracy trade-off. While earlier architectures like RNN-T and Transformer Transducers focused on minimizing delay \cite{graves2012rnnt, zhang2020transformerTransducer}, recent large-scale models such as Whisper have prioritized robustness across diverse languages \cite{radford2023whisper}. However, this shift often introduces ambiguity in continuous-script languages like Thai, where the lack of explicit word boundaries means that segmentation errors can propagate downstream, altering the pragmatic meaning of a user's intent \cite{chormai2020thai}.

Once a textual representation is established, RAG has become the standard for grounding responses in factual knowledge \cite{lewis2020rag}. This paradigm has rapidly expanded to include LLMs that utilize external APIs and tools, effectively functioning as interactive agents \cite{xi2023llmAgents, schick2023toolformer}. Frameworks such as ReAct explicitly interleave chain-of-thought (CoT) reasoning with external actions \cite{yao2023react}. While these methods enhance the system's reasoning capabilities, the temporal coordination of these asynchronous external operations remains a significant, yet under-analyzed, interactional challenge.

At the output layer, end-to-end neural TTS models, such as VITS and Tacotron 2, have enabled highly naturalistic synthesis \cite{kim2021vits, shen2018tacotron2}. Recent advancements have focused on cross-lingual voice cloning and expressive consistency to maintain a stable agent persona \cite{gururani2019prosody, zhang2023seamless}. While prior work has largely optimized these components in isolation, the interactional consequences of their coupled behavior remain under-analyzed. Our work addresses this gap by treating the speech pipeline not as a series of models, but as a complete interactional infrastructure where temporal and semantic seams directly govern user experience.

\section{System Architecture}~\label{system}

\begin{figure*}[t]
    \centering
    \includegraphics[width=\textwidth]{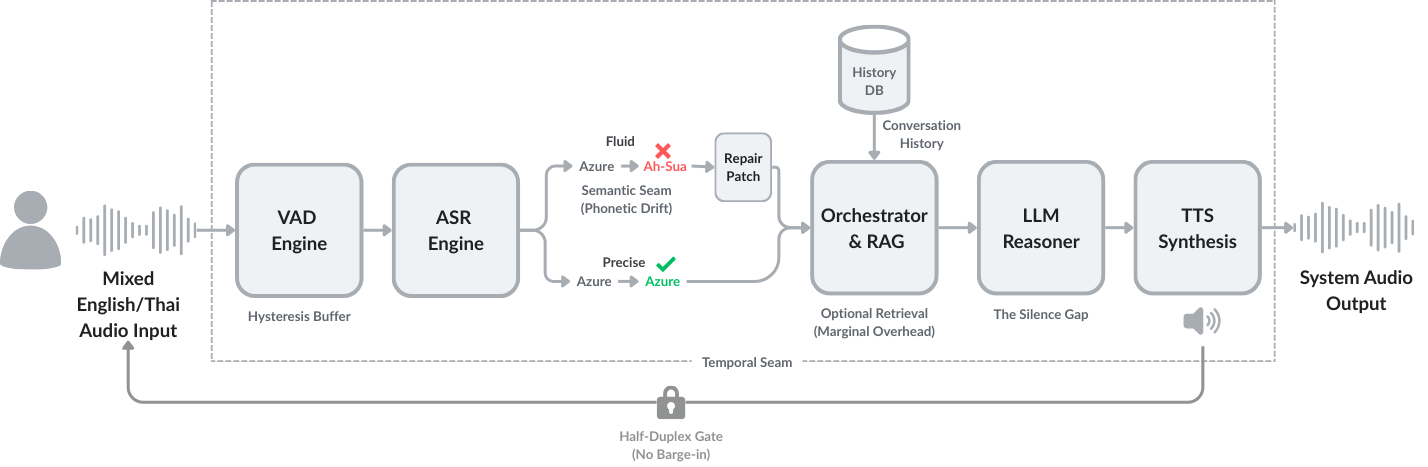} 
    \caption{\textbf{Interactional Friction in an S2S-RAG Architecture.}}
    \label{fig:architecture}
\end{figure*}

The system analyzed in this paper is a modular, event-driven pipeline designed for real-time, speech-to-speech interaction. Its architecture is representative of common production deployments that must balance responsiveness with the control offered by specialized components. The following subsections describe the end-to-end data flow, detailing each layer from the initial audio signal processing to the final synthesized speech output. A high-level schematic of this architecture is provided in Figure~\ref{fig:architecture}.

\subsection{The Hearing Layer: Signal Stabilization and Floor Management}~\label{hearing_layer}
The interaction entry point is a high-performance bidirectional streaming service. However, raw audio ingestion is insufficient for natural interaction; the system must algorithmically determine when a user has effectively ``yielded the floor.'' We implemented a custom Voice Activity Detection (VAD) that transforms acoustic energy into interactional state changes. To prevent the system from interrupting users during natural mid-sentence pauses (``micro-pauses"), we engineered a stabilization layer using two specific signal processing techniques:

\begin{itemize}
    \item \textbf{Input Normalization:} The system first standardizes variable client-side audio streams into a uniform spectral representation. This ensures that the detection logic remains robust against hardware variations and background noise.
    \item \textbf{Probabilistic Hysteresis:} Raw detection probabilities are volatile. We implemented a temporal smoothing algorithm combined with asymmetric thresholds. The system demands high confidence to recognize a ``Turn Start'' but allows lower confidence to sustain a turn. This hysteresis effectively ``holds the floor'' for the user, deliberately sacrificing milliseconds of latency to prevent the rudeness of premature interruption.
\end{itemize}

Once the system determines that the user’s turn has ended, the audio payload is encapsulated and transferred via an asynchronous message queue.

\subsection{The Orchestrator: Concurrency and Contextual Latency}
The backend orchestration is handled by a high-concurrency service responsible for maintaining the illusion of a coherent memory. This component acts as the ``interactional router,'' mediating between the real-time constraints of speech and the asynchronous nature of reasoning. This layer enforces a critical path that defines the system's ``Time-to-Reasoning'':

\begin{itemize}
    \item \textbf{Asynchronous Concurrency Control:} To maintain stability under load, the system utilizes a semaphore-based locking mechanism. From an HCI perspective, this introduces a hidden queue. During high traffic, users do not experience a failure, but an extended silence that mimics cognitive hesitation.
    \item \textbf{Stateful Context Injection:} Before the AI model is invoked, the orchestrator retrieves the full conversation history from a persistent data store. This engineering choice prioritizes interactional grounding—ensuring the agent knows what was previously said—over immediate responsiveness. 
\end{itemize}

This architecture reveals that the ``unnatural'' pauses in S2S-RAG systems are not accidental; they are the accumulated structural cost of transforming a raw signal into a context-aware response.

\subsection{The Reasoning Core: LLM and RAG}
Once the Orchestrator assembles the complete prompt---containing the user's transcribed utterance and the relevant conversation history---it is passed to the system's cognitive core. This layer is responsible for understanding the user's intent, retrieving relevant knowledge, and generating a coherent textual response.

The primary component of this layer is a LLM. The LLM's function is to perform the complex reasoning required to formulate an answer. The selection of the LLM class (e.g., a lightweight model for speed vs. a high-fidelity model for accuracy) is a critical architectural decision that directly creates the performance trade-offs quantified in Section~\ref{llm_trade_off}.

To ensure responses are grounded in factual and proprietary data, we implement a RAG strategy. Before invoking the LLM, the Orchestrator queries a vector database to retrieve relevant document chunks based on the user's query. These chunks are then injected directly into the LLM's context window. As noted in Figure~\ref{fig:architecture}, this retrieval step adds a marginal, constant overhead but is essential for controlling hallucinations and providing auditable, data-driven answers. The final output of this layer is a complete, text-only response, which is then handed off to the Speaking Layer for synthesis.

\subsection{The Speaking Layer: Synthesis and Interactional Deafness}

The final stage of the modular pipeline is the Speaking Layer, which is responsible for converting the LLM's text-based response into system audio output (Figure~\ref{fig:architecture}). This is handled by a modern neural TTS engine, such as those discussed in Section~\ref{background}. While these models can achieve highly naturalistic synthesis, this architectural stage introduces two primary forms of interactional friction.

First, the modular conversion from text back to audio often results in what we term Expressive Flattening. The rich, pragmatic intent embedded in the LLM's text (e.g., certainty, hesitation) is frequently lost, leading to a voice that, while clear, may sound monotonous or emotionally disconnected from the context of the conversation.

Second, to manage turn-taking, our architecture implements a rigid Half-Duplex Gate. This mechanism creates an exclusive lock on the audio output channel the moment TTS playback begins, making the system effectively ``deaf'' to any user input. This architectural choice prevents users from correcting the system or providing feedback via ``barge-in''. As analyzed in Section~\ref{half_duplex}, this forced-listening state is a primary source of user frustration, turning simple repairs into lengthy, multi-turn error loops.

\section{Architectural Trade-offs: System-Level Performance}~\label{pipeline}

To quantify the interactional friction in our modular pipeline, we evaluated the system using a custom dataset of customer support dialogues. We utilized an ``LLM-as-a-Judge'' methodology to calculate Normalized Word Error Rate (WER) and Semantic Correction Scores.

\textit{Experimental Scope:} The results below isolate the neural processing time (ASR + LLM + TTS). While the full architecture supports RAG, our implementation utilizes an optimized fast local retrieval strategy. Therefore, these metrics represent the baseline interactional latency, as the RAG process adds only a negligible, constant overhead.

Crucially, we isolated code-switching---the alternation between local language and English technical terms---as our primary stress test. In enterprise environments, users rarely speak a pure language; they intersperse domain-specific English jargon (e.g., ``Deploy,'' ``Cloud,'' ``Microservice'') within local sentences. This represents a critical challenge for modular architectures: if the ASR produces phonetic transliterations, it causes a cascading failure where the downstream LLM receives meaningless text.

\subsection{The ASR Trade-off: Speed vs. Robustness}~\label{asr_trade_off}
The choice of ASR engine establishes the fundamental conversational rhythm. Our benchmarks reveal a critical trade-off between latency and robustness to technical jargon.

\begin{itemize}
    \item \textbf{Typhoon, a Thai ASR model:} Optimized for speed, this model achieved an ultra-low latency of \textbf{417.1 ms}. However, it exhibited a high Normalized WER of 0.562 and failed the code-switching stress test. For example, the term ``Azure'' was transcribed into phonetic nonsense, causing the LLM to fail. This highlights its unsuitability for technical domains where jargon is common.
    \item \textbf{Google Cloud STT V1:} While significantly slower at \textbf{2457.2 ms}, this engine achieved a much lower Normalized WER of 0.243. Its superior accuracy enabled it to pass the stress test by correctly resolving jargon like ``PostgreSQL'' and ``Bandwidth.'' This confirms that for professional applications, higher latency is the current price of achieving semantic reliability.
\end{itemize}

\subsection{The LLM Trade-off: Latency vs. Reasoning}~\label{llm_trade_off}
In a modular pipeline, algorithmic ``intelligence'' is a direct temporal cost that varies with the model's computational depth. We benchmarked two model classes to quantify this trade-off.

\begin{itemize}
    \item \textbf{Lightweight Models (e.g., Gemini 2.5 Flash):} Optimized for speed, these models average \textbf{1148.6 ms} in latency. They enable a fluid conversational experience but risk failing on complex, multi-step tasks.

    \item \textbf{High-Fidelity Models (e.g., GPT-5):} Engineered for accuracy, their latency for complex queries spikes to \textbf{5264.1 ms}. This temporal cost reflects deeper chain-of-thought processing.

    \item \textbf{The Architectural Dilemma:} This latency difference forces a choice between a fast model that risks errors and a slow one that creates an untenable silence gap. When combined with ASR delay, a high-fidelity model produces a pause exceeding eight seconds, making mitigation techniques (e.g., filler audio) non-negotiable for voice interfaces.
\end{itemize}

\subsection{The Hybrid Solution: A Textual Repair Layer}
To resolve the ASR trade-off, we engineered a hybrid solution: an intermediate Textual Repair Layer. Powered by a lightweight LLM (e.g., Gemini 2.5 Flash-Lite), this component is inserted between the fast ASR and the primary reasoning LLM to act as a real-time ``post-edit'' filter.

\begin{itemize}
   \item \textbf{Functionality and Performance:} The layer specifically targets code-switching failures. As measured by our LLM-as-a-Judge methodology, it achieved a high efficacy with a Semantic Correction Score of \textbf{0.85}, successfully re-mapping phonetic transliterations (e.g., restoring the Thai phonetic spelling of ``AWS'' to its English acronym).

    \item \textbf{Latency Cost:} This high-accuracy correction is achieved with a minimal and fixed latency overhead of only \textbf{623 ms}.

    \item \textbf{Superior Latency Profile:} The resulting hybrid ``Fluid Pipeline'' (Fast ASR + Repair) has a total latency of \textbf{1040 ms}. This is less than half of the \textbf{$\sim$2457 ms} required by the ``Precise Pipeline'' that waits for a slower ASR, proving it is more efficient to correct an imperfect stream than to wait for a batch-accurate model.
\end{itemize}

\subsection{The End-to-End Comparison: Cost, Latency, and Control}
The final pipeline component, TTS synthesis, adds a consistent overhead of $\sim$450 ms. Aggregating all components, we compared our modular tiers against the industry benchmark, GPT-Realtime (Table \ref{tab:pipeline_summary}), revealing a divergence between cost, control, and native prosody.

\begin{itemize}
    \item \textbf{Cost Efficiency:} Our ``Fluid'' pipeline (\$0.0010/turn) is approximately \textbf{15x cheaper} than GPT-Realtime (\$0.0154/turn). For high-volume enterprise applications, this cost difference is architecturally decisive.
    \item \textbf{Latency Profile:} The modular Fluid pipeline (2--3s) is faster than GPT-Realtime (4--6s). This is likely because our text-based intermediate steps are computationally lighter than GPT-Realtime's dense audio-to-audio processing.
    \item \textbf{The ``Black Box'' Problem:} While GPT-Realtime offers superior emotional prosody, it is a black box. A developer cannot fix recognition errors for company-specific terms (e.g., ``AXONS''). In contrast, our modular \textit{Precise} tier allows explicit injection of ``Phrase Sets'' into the ASR, guaranteeing domain accuracy at the cost of prosodic naturalness.
\end{itemize}

\begin{table}[h]
\centering
\small
\caption{End-to-End Pipeline Tiers vs. Industry Benchmark}
\label{tab:pipeline_summary}
\begin{tabular}{|p{0.25\columnwidth}|p{0.15\columnwidth}|p{0.45\columnwidth}|}
\hline
\textbf{Pipeline} & \textbf{Cost/Turn} & \textbf{Total Latency \& Use Case} \\
\hline
Fluid (Typhoon) 
& \$0.0010 
& 2--3 s. Best for conversational chit-chat. \\
\hline
Precise (Google) 
& \$0.0023 
& 3--5 s. Best for FAQ and strict terminology. \\
\hline
Reasoning (GPT-5) 
& \$0.0046 
& 5--7 s. Best for complex planning. \\
\hline
GPT-Realtime
& \$0.0154
& 4--6 s. Benchmark (Black box). \\
\hline
\end{tabular}
\end{table}
\section{Interaction Challenges: User-Facing Consequences}~\label{Challenges}

The performance metrics quantified in Section~\ref{pipeline}---sub-second ASR latencies, multi-second reasoning gaps, and accuracy trade-offs---are not merely engineering benchmarks. As our analysis reveals, they translate directly into user-facing conversational friction that fundamentally alters the pragmatics of the interaction. We categorize these consequences into four key interactional breakdowns.

\subsection{The Deafness of the ``Half-Duplex'' Gate}~\label{half_duplex}
Human conversation is inherently collaborative; listeners backchannel (``mm-hmm'') and speakers yield to interruptions. However, our architectural decision to use a Half-Duplex Gate (Section~\ref{hearing_layer}) creates a rigid ``turn-taking collision.'' Because the system creates an exclusive lock on the audio channel while speaking, it becomes effectively deaf to user interruptions.
\begin{itemize}
    \item \textbf{The Scenario:} A user notices the ASR has mis-transcribed ``Azure'' as nonsense (the code-switching error identified in Section~\ref{asr_trade_off}).
    \item \textbf{The Friction:} The user attempts to ``barge in'' to correct the error immediately. However, the system continues to process the hallucinated intent and generates a full response based on the wrong premise.
    \item \textbf{Result:} The user is forced to wait for the system to finish its mistake before they can issue a correction. This turns a 2-second repair into a 15-second loop of frustration. This breakdown in conversational repair, which we term Repair Rigidity, is a direct consequence of the system's inability to listen while speaking.
\end{itemize}

\textbf{Design Implication:} Future modular systems must implement Acoustic Echo Cancellation (AEC) to enable an ``Always-Listening'' state, allowing the user's voice to trigger a hardware interrupt that halts generation immediately.

\subsection{The Cost of Literalness: Expressive Flattening}
A key source of friction arises from what we term Expressive Flattening. In one test scenario, a user, frustrated with a repeated error, asked with heavy sarcasm, \textit{``Great, so you're telling me it failed again?''}. In a human conversation, the sarcastic tone signals frustration. However, after being transcribed to plain text, the LLM interpreted the word ``Great'' literally and responded cheerfully: \textit{``I'm glad to hear that! How else can I assist you?''}. This wildly inappropriate response, caused by the loss of paralinguistic cues (prosody, pitch, cadence) during the ASR conversion, instantly broke the user's trust and escalated their frustration. This demonstrates that for modular systems, the semantic meaning of an utterance is often only half the story.

\textbf{Design Implication:} Future conversational agents cannot rely on text alone. They must incorporate models that can either infer emotional prosody from the raw audio signal or, at a minimum, recognize uncertainty. When a literal interpretation starkly contrasts with the conversational context (e.g., saying ``Great'' after a failure), the system should be designed to seek clarification (e.g., ``I'm sorry, I might have misunderstood. Could you clarify?'') rather than generating a naive, literal response.

\subsection{The Trade-Off Between Responsiveness and Semantic Integrity}
Our experiments with Typhoon ASR (417 ms) revealed a counter-intuitive finding: \textit{speed can be detrimental if it lacks semantic certainty.}
When the system responds in under 500 ms but creates a ``Semantic Drift'' error (misinterpreting English jargon), the interaction feels ``manic'' rather than responsive. The user perceives the agent as jumping to conclusions. Conversely, the Google STT pipeline (2.4s latency) feels ponderous, but its high accuracy fosters a sense of deliberation.

\textbf{Design Implication:} Latency should be dynamic. Users forgive delays if they perceive the task as complex. A system should respond instantly to ``Hello'' (Phatic) but deliberately pause (or use filler words like ``Let me check...'') for complex technical queries (Epistemic), aligning the wait time with the user's mental model of the task difficulty.

\subsection{The Pragmatic Failure of Silence}
In the Deep Reasoning pipeline (GPT-5), the total Turn Delay reaches 8-9 seconds. This significant delay is a primary example of Temporal Misalignment, where the system's response time violates the user's conversational expectations.
 In a text interface, this is a ``loading state.'' In a voice interface, silence is a social signal that usually implies hesitation, confusion, or a dropped connection.
Without visual feedback, users in our tests frequently asked, ``Are you still there?'' after just 3 seconds of silence. This verbal checking resets the VAD, inadvertently cancelling the pending request and restarting the loop.

\textbf{Design Implication:} The system must ``hold the floor'' explicitly. Instead of silence, the Orchestrator should inject ``Audio Spinners''---non-lexical fillers or breath sounds---to signal that the agent is thinking, not broken.

\subsection{The Paradox of Control vs. Fluidity}
Finally, our comparison with GPT-Realtime highlights the fundamental trade-off of the S2S-RAG architecture.
\begin{itemize}
  \item Native Models (GPT-Realtime) offer superior prosody and seamless turn-taking but act as a ``Black Box.'' If the model fails to recognize a proprietary term like ``AXONS,'' the developer has no mechanism to fix it.
  \item Modular Pipelines (Our Approach) are clunkier (higher latency, flatter voice) but offer Determinism. We can inject specific ``Phrase Sets'' into the Google ASR and use repair patch layers to hard-code fixes for domain jargon.
\end{itemize}
For enterprise applications, this means the ``Interactional Friction'' of the modular pipeline is currently a necessary cost. The ability to guarantee the accuracy of business terminology (via the discrete ASR/LLM seams) outweighs the aesthetic benefits of the smoother, but uncontrollable, black-box models.

\textbf{Design Implication:} The future of enterprise voice AI lies in \textit{Hybrid Routing}. Rather than choosing a single architecture, the Orchestrator should dynamically switch between models based on intent. It should utilize a fluid end-to-end model for casual, social turns (phatic communication) to build rapport, but seamlessly switch to a modular, deterministic pipeline when high-stakes business entities or data retrieval are required.

\subsection{The Token-to-Audio Mismatch: The ``Stutter'' of Streaming}
While LLMs stream text token-by-token (``typewriter effect''), TTS engines require coherent linguistic units (phrases or sentences) to generate natural prosody. This creates a synchronization friction we term \textit{The Stream Aggregation Bottleneck}.

\begin{itemize}
  \item \textbf{The Scenario:} The pipeline attempts to minimize latency by sending LLM output to the TTS engine as soon as possible.
  \item \textbf{The Friction:} In English, the system might naively split by whitespace. However, in continuous-script languages like Thai, there are no explicit word boundaries. If the system sends a partial chunk to the TTS (e.g., splitting a compound Thai word in half), the TTS may mispronounce the tones or produce a robotic, staccato rhythm.
  \item \textbf{Result:} The system faces a ``Jitter Trade-off.'' If it buffers too little text, the voice sounds choppy and incoherent (``Audio Glitching''). If it buffers too much (waiting for a full sentence delimiter), it re-introduces the very latency the streaming architecture tried to remove. This results in an interaction that feels visually fast (on screen) but audibly sluggish.
\end{itemize}

\textbf{Design Implication:} S2S pipelines require a dedicated \textit{Heuristic Aggregator} between the LLM and TTS. For languages like Thai, this algorithm cannot rely on simple delimiters. It must perform real-time syntactic analysis to identify ``breath groups'' or semantic phrases, ensuring the TTS receives a chunk that is long enough to carry proper intonation but short enough to maintain low latency.
\section{Conclusion}~\label{conclusion}

Achieving fluid conversational AI is not a race to zero latency but a challenge of infrastructure design. This paper has demonstrated that in modular S2S-RAG systems, ``interactional friction''---manifesting as unnatural silence, semantic errors, or interactional deafness---is an unavoidable structural cost paid for the control and economic viability that enterprise applications demand.

Our analysis reveals that the engineering trade-offs between speed, accuracy, and cost are not abstract benchmarks; they are directly experienced by users as the felt quality of the conversation. This reframes the work of the AI engineer from merely optimizing isolated models to carefully choreographing the temporal and semantic seams of the entire pipeline. Until monolithic end-to-end models can offer the granular control and cost-efficiency of modular architectures, the art of building natural speech interfaces will lie not in eliminating latency, but in skillfully turning a distributed system into a coherent conversational partner.
\section{Future Work}

While this paper focused on the one-to-one interactional friction within a half-duplex S2S-RAG pipeline, our findings lay the groundwork for several critical areas of future research.

\subsection{Multi-Party Conversation and Floor Management}
Our analysis was confined to a dyadic (two-party) interaction. The next frontier is the multi-party or group conversation scenario. This introduces exponential complexity, requiring the system to not only manage its own turn-taking but also to perform speaker diarization (i.e., identifying who is speaking), handle cross-talk and overlapping speech, and determine when the ``floor'' has been yielded by the entire group. The rigid Half-Duplex Gate, a source of friction in our current work, would be entirely unworkable in such a dynamic environment.

\subsection{Enabling Full-Duplex Interaction via Incremental Processing}
A key implication of our work is the need to eliminate the ``interactional deafness'' caused by half-duplex gating. Future work must focus on implementing true full-duplex systems. This requires not only robust AEC to allow for ``barge-in,'' but also a shift towards fully incremental ASR, LLM, and TTS components. An incremental pipeline would allow the system to process user speech, reason, and begin generating a response in real-time, and crucially, be interrupted and re-plan mid-utterance.

\subsection{Proactive and Expressive Latency Management} 
Our paper suggests using audio fillers to manage the ``Silence Gap,'' but this can be significantly improved. A more advanced approach would reframe latency as a resource for conversational grounding. Rather than deploying generic phrases, future systems could learn to use the reasoning delay to confirm their interpretation of the task (e.g., ``Understood. You need the deployment logs for the production server from yesterday...''). This strategy makes the wait-time feel productive. Furthermore, prosodic control could be leveraged to generate subtle, human-like non-lexical vocalizations---such as a thoughtful hum or a soft intake of breath---to signal that the system is processing information, turning a potentially awkward silence into a shared, collaborative moment.

\subsection{Multimodal Signal Integration}
Finally, this work was limited to audio signals. Future research should integrate multimodal signals---such as gaze, head pose, and gestures---into the interactional floor management. A user breaking eye contact might signal the end of their turn, while a furrowed brow could trigger a clarification request from the agent without a single word being spoken. These additional channels would provide the rich, parallel streams of information that humans use to seamlessly choreograph conversations, helping to finally smooth the seams of modular AI systems.


\bibliographystyle{ACM-Reference-Format}
\bibliography{sample-base}

@String{Computer = "{IEEE} Computer" }

@book{schegloff2007sequence,
  title     = {Sequence Organization in Interaction: A Primer in Conversation Analysis},
  author    = {Schegloff, Emanuel A.},
  year      = {2007},
  publisher = {Cambridge University Press}
}

@article{skantze2021survey,
  title   = {Turn-Taking in Conversational Systems and Human--Robot Interaction: A Review},
  author  = {Skantze, Gabriel},
  journal = {Computer Speech and Language},
  volume  = {67},
  pages   = {101178},
  year    = {2021}
}

@article{schlangen2011incremental,
  title   = {A General, Abstract Model of Incremental Dialogue Processing},
  author  = {Schlangen, David and Skantze, Gabriel},
  journal = {Dialogue \& Discourse},
  volume  = {2},
  number  = {1},
  pages   = {83--111},
  year    = {2011}
}

@inproceedings{graves2012rnnt,
  title     = {Sequence Transduction with Recurrent Neural Networks},
  author    = {Graves, Alex},
  booktitle = {Proceedings of the ICML Workshop on Representation Learning},
  year      = {2012}
}

@inproceedings{zhang2020transformerTransducer,
  title     = {Transformer Transducer: A Streamable Speech Recognition Model},
  author    = {Zhang, Yanzhang and Qin, Jiahui and Zhang, Daniel and others},
  booktitle = {Proceedings of ICASSP},
  year      = {2020}
}

@inproceedings{radford2023whisper,
  title     = {Robust Speech Recognition via Large-Scale Weak Supervision},
  author    = {Radford, Alec and Kim, Jong Wook and Xu, Tao and Brockman, Greg and McLeavey, Christine and Sutskever, Ilya},
  booktitle = {Proceedings of the 40th International Conference on Machine Learning (ICML)},
  year      = {2023}
}

@inproceedings{chormai2020thai,
  title     = {Neural Thai Word Segmentation},
  author    = {Chormai, Pucktada and others},
  booktitle = {Proceedings of a Workshop on Asian Language Processing},
  year      = {2020},
  note      = {On neural approaches to Thai word segmentation}
}

@inproceedings{lewis2020rag,
  title     = {Retrieval-Augmented Generation for Knowledge-Intensive {NLP} Tasks},
  author    = {Lewis, Patrick and Perez, Ethan and Piktus, Aleksandra and Petroni, Fabio and Karpukhin, Vladimir and Goyal, Naman and K{\"u}ttler, Heinrich and Lewis, Mike and Yih, Wen-tau and Rockt{\"a}schel, Tim and Riedel, Sebastian and Kiela, Douwe},
  booktitle = {Advances in Neural Information Processing Systems (NeurIPS)},
  year      = {2020}
}

@inproceedings{schick2023toolformer,
  title     = {Toolformer: Language Models Can Teach Themselves to Use Tools},
  author    = {Schick, Timo and Dwivedi-Yu, Jane and Dess{\`\i}, Roberto and Raileanu, Roberta and Lomeli, Maria and Hambro, Eric and Zettlemoyer, Luke and Cancedda, Nicola and Scialom, Thomas},
  booktitle = {Advances in Neural Information Processing Systems (NeurIPS)},
  year      = {2023}
}

@article{xi2023llmAgents,
  title   = {The Rise and Potential of {LLM}-Based Agents: A Survey},
  author  = {Xi, Xiaoyu and others},
  journal = {arXiv preprint arXiv:2309.07864},
  year    = {2023}
}

@inproceedings{yao2023react,
  title     = {ReAct: Synergizing Reasoning and Acting in Language Models},
  author    = {Yao, Shunyu and Zhao, Jeffrey and Yu, Dian and Du, Nan and Shafran, Izhak and Narasimhan, Karthik and Cao, Yuan},
  booktitle = {Proceedings of the International Conference on Learning Representations (ICLR)},
  year      = {2023}
}

@inproceedings{shen2018tacotron2,
  title     = {Natural {TTS} Synthesis by Conditioning WaveNet on Mel Spectrogram Predictions},
  author    = {Shen, Jonathan and Pang, Ruoming and Weiss, Ron J. and others},
  booktitle = {Proceedings of ICASSP},
  year      = {2018}
}

@inproceedings{kim2021vits,
  title     = {{VITS}: Conditional Variational Autoencoder with Adversarial Learning for End-to-End Text-to-Speech},
  author    = {Kim, Jaehyeon and Kong, Jungil and Son, Juhee},
  booktitle = {Proceedings of ICML},
  year      = {2021}
}

@article{gururani2019prosody,
  title   = {Prosody Transfer in Neural Text to Speech Using Global Pitch and Loudness Features},
  author  = {Gururani, Siddharth and Gupta, Kilol and Shah, Dhaval and Shakeri, Zahra and Pinto, Jervis},
  journal = {arXiv preprint arXiv:1911.09645},
  year    = {2019}
}

@article{zhang2023seamless,
  title   = {Seamless{M4T}: Massively Multilingual and Multimodal Machine Translation},
  author  = {{Seamless Communication} and others},
  journal = {arXiv preprint arXiv:2308.11596},
  year    = {2023}
}

\appendix

\end{document}